\documentclass[showpacs,preprintnumbers,superscriptaddress]{revtex4}
\usepackage{CJK}
\usepackage{amsmath,amssymb,graphicx,bm}
\begin{document}

\title{Accretion onto a Kiselev black hole}

\author{Lei Jiao}
\affiliation{College of Physical Science and Technology, Hebei University, Baoding 071002, China}
\author{Rongjia Yang \footnote{Corresponding author}}
\email{yangrongjia@tsinghua.org.cn}
\affiliation{College of Physical Science and Technology, Hebei University, Baoding 071002, China}
\affiliation{Hebei Key Lab of Optic-Electronic Information and Materials, Hebei University, Baoding 071002, China}

\begin{abstract}
We consider accretion onto a Kiselev black hole. We obtain the fundamental equations for accretion without the back-reaction. We determine the general analytic expressions for the critical points and the mass accretion rate and find the physical conditions the critical points should fulfill. The case of polytropic gas are discussed in detail. It turns out that the quintessence parameter plays an important role in the accretion process.
\end{abstract}


\maketitle

\section{Introduction}
A number of astronomical observations confirmed that the universe is undergoing an accelerated expansion. In order to explain this phenomenon, an unknown energy component, dubbed dark energy, must have been introduced in the framework of general relativity. The simplest candidate is the cosmological constant model which is consistent with most of the current astronomical observations; however, it suffers from the cosmological constant problem \cite{Weinberg:1988cp, Carroll:2000fy, Bull:2015stt} and maybe the age problem \cite{Yang:2009ae} as well. It is thus natural to consider other complicated cases. A dynamic scalar field can also serve as the dark energy component such as quintessence, phantom, k-essence, etc. Quintessence is the simplest scalar field dark energy model without having theoretical problems like Laplacian instabilities or ghosts. The energy density and the pressure of quintessence vary with time depending on the scalar field and the potential, which are, respectively, given by $\rho =\frac{1}{2}\dot{\phi}^2+V(\phi)$ and $p =\frac{1}{2}\dot{\phi}^2-V(\phi)$. One Schwarzschild-like solution related to the quintessence model was found in \cite{Kiselev:2002dx}. The solution describes a spherically symmetric and static exterior spacetime surrounded by a quintessence field. The gravitational lensing due to this Schwarzschild-like black hole was investigated in \cite{Younas:2015sva}. Here, we consider accretion of matter onto this type black hole. We will discuss the critical radius, critical fluid velocity, the speed of sound, the mass accretion rate, and so on.

Accretion of matter onto black hole is an important phenomenon of long-standing interest to astrophysicists. Pressure-free gas being dragged onto a massive central object was examined in \cite{hoyle1939effect,Bondi:1944jm}. This study was generalized to the case of the spherically symmetric and transonic accretion of adiabatic fluids onto astrophysical objects \cite{Bondi:1952ni}. The steady-state spherical symmetric flow of matter into or out of a condensed object in the framework of general relativity was investigated in \cite{michel1972accretion}. Issues of the critical points of accretion were discussed in \cite{begelman1978accretion, Malec:1999dd}.
Accretion has been analyzed in various cases in the literature, including onto a charged black hole \citep{michel1972accretion,Jamil:2008bc}, onto a Kerr-Newman black hole \citep{Babichev:2008dy, JimenezMadrid:2005rk,Bhadra:2011me}, onto a Reissner-Nordstrom black hole \citep{Babichev:2008jb}, onto a moving black hole \citep{petrich1988accretion}, onto a higher dimensional black hole \citep{Giddings:2008gr,Sharif:2011ih,John:2013bqa,Debnath:2015yva}, onto a black hole in a string cloud background \citep{Ganguly:2014cqa}, and onto cosmological black holes or onto Schwarzschild-(anti-)de Sitter spacetimes \citep{Mach:2013fsa, Mach:2013gia, Karkowski:2012vt}. Quantum gravity corrections to accretion onto a Schwarzschild black hole were considered in the context of asymptotically safe scenario \cite{Yang:2015sfa}. Accretion of phantom energy onto a black hole may decrease the black-hole mass \citep{Babichev:2004yx, Babichev:2005py}, however, it may not be the case if using solutions describing black holes in a Friedmann-Robertson-Walker universe \citep{Gao:2008jv}. In \cite{Liu:2009ts}, an exact solution was derived for one or two dust shells collapsing towards a pre-existing black hole.

The rest of the paper is organized as follows. In the next section, we will present the fundamental equations for matter accreting on to a Kiselev black hole. In Sect. III, we will determine the critical points and the conditions they must fulfill. In Sect. IV, as an application, we will discuss the case of polytropic gas in detail. Finally, we will briefly summarize and discuss our results in Sect. V.

\section{Basic equations for accretion}
We consider a static and spherically symmetric black hole surrounded by the quintessence, its geometry is given by \cite{Kiselev:2002dx}
\begin{eqnarray}
\label{Sp}
ds^2=-f(r)dt^{2}+f^{-1}(r)dr^{2}+r^{2}(d\theta^{2}+\sin^{2}\theta d\varphi^{2}),
\end{eqnarray}
where
\begin{eqnarray}
f(r)=1-\frac{2M}{r}-\frac{\sigma}{r^{3w+1}}
\end{eqnarray}
where $w=p/\rho$ is the equation of state of quintessence, $\sigma$ is the quintessence parameter, and $M$ is the mass of the black hole measured by an observer at infinity. Here we take the units $G=c=1$ and focus on the special case $w=-2/3$, which describes a Schwarzschild-like black hole surrounded by quintessence
\begin{eqnarray}
f(r)=1-\frac{2M}{r}-\sigma r.
\end{eqnarray}
The horizons can be obtained by solving $g_{00}=0$ and are found to be: $r_{\pm}=\frac{1\pm\sqrt{1-8M\sigma}}{2\sigma}$ with $0<\sigma\leq\frac{1}{8M}$. The region $r=r_{+}$ corresponds to the outer horizon while $r=r_{-}$ represents the inner horizon. For $\sigma=\frac{1}{8M}$, it is a Schwarzschild black hole with only one horizon, $r_{\rm h}=8M$.

We suggest the steady-state radial inflow matter onto the Schwarzschild-like without back-reaction and treat the quintessence as the background field. The matter is approximated as a perfect fluid specified by the energy-momentum tensor
\begin{eqnarray}
\label{em}
T_{\alpha\beta}=(\rho+p)u_{\alpha}u_{\beta}+pg_{\alpha\beta}
\end{eqnarray}
where $\rho$ and $p$ are the proper energy density and the proper pressure of the fluid, respectively. The fluid 4-velocity $u^{\alpha}=dx^{\alpha}/ds$ obeys the normalization condition $u^{\mu}u_{\mu}=-1$. The radial component of the 4-velocity is denoted as $u=dr/ds$. For $\mu>1$ the components of velocity vanish, one can easily get
\begin{eqnarray}
\label{con1}
(u^{0})^{2}=\frac{1-\frac{2M}{r}-\sigma r+u^{2}}{\left(1-\frac{2M}{r}-\sigma r\right)^{2}}.
\end{eqnarray}
In the local inertial rest frame of the fluid, it is also usual to define the proper baryon number density $n$ and the baryon number flux $J^{\mu}=nu^{\mu}$. The conservation of mass-energy for an adiabatic fluid is given by
\begin{eqnarray}
\label{cme}
0=Tds=d\left(\frac{\rho}{n}\right)+pd\left(\frac{1}{n}\right),
\end{eqnarray}
from which we can obtain
\begin{eqnarray}
\label{cme1}
\frac{d\rho}{dn}=\frac{\rho+p}{n}.
\end{eqnarray}
The adiabatic sound speed of the fluid is defined as
\begin{eqnarray}
\label{ss}
a^2 \equiv\frac{dp}{d\rho}=\frac{n}{\rho+p}\frac{dp}{dn}.
\end{eqnarray}
If no particle are created or destroyed, the conservation of particle number gives
\begin{eqnarray}
\label{con1}
J^{\mu}_{~;\mu}=(nu^{\mu})_{;\mu}=0,
\end{eqnarray}
where $;$ denotes the covariant derivative with respect to the coordinate. For the Schwarzschild-like metric, Eq. (\ref{con1}) can be rewritten as
\begin{eqnarray}
\label{con1s}
\frac{1}{r^{2}}\frac{d}{dr}(r^2nu)=0,
\end{eqnarray}
for a perfect fluid it gives the integration as
\begin{eqnarray}
\label{con1s1}
r^2nu=C_1,
\end{eqnarray}
where $C_1$ is an integration constant. Integrating equation (\ref{con1s}) over the spatial volume and multiplying by the mass of each particle, $m$, we derive
\begin{eqnarray}
\label{acc}
\dot{M}=4\pi r^2mnu.
\end{eqnarray}
where $m$ is the mass of the particle of the gas and $\dot{M}$ is a constant of integration having dimensions of mass per unit time. In fact, this is the Bondi mass accretion rate.

Assuming the accretion of adiabatic fluids does not disturb the global spherical symmetry of the black hole, the $\nu=0$ component of the energy-momentum conservation $T^{\mu}_{~\nu;\mu}=0$ gives
\begin{eqnarray}
\label{con2s}
\frac{1}{r^{2}}\frac{d}{dr}\left[r^2(\rho+p)u\left(1-\frac{2M}{r}-\sigma r+u^2\right)^{1/2}\right]=0,
\end{eqnarray}
which can be integrated
\begin{eqnarray}
\label{con2s1}
r^2(\rho+p)u\left(1-\frac{2M}{r}-\sigma r+u^2\right)^{1/2}=C_2,
\end{eqnarray}
where $C_2$ is an integration constant. Dividing Eqs. (\ref{con2s1}) and (\ref{con1s1}) and then squaring, we obtain
\begin{eqnarray}
\label{con3}
\left(\frac{\rho+p}{n}\right)^2\left[1-\frac{2M}{r}-\sigma r+u^2\right]=\left(\frac{\rho_\infty+p_\infty}{n_\infty}\right)^2.
\end{eqnarray}
The $\nu=1$ component of the energy-momentum conservation $T^{\mu}_{~\nu;\mu}=0$ gives
\begin{eqnarray}
\label{con11}
u\frac{du}{dr}=-\frac{1-\frac{2M}{r}-\sigma r+u^2}{\rho+p}\frac{dp}{dr}-\frac{1}{2}\left(\frac{2M}{r^{2}}-\sigma\right)
\end{eqnarray}
Eqs. (\ref{acc}) and (\ref{con3}) are fundamental equations for the flow of matter onto the Schwarzschild-like black hole with the back-reaction of matter ignored.

\section{Conditions for critical accretion}
In this section, we discuss the conditions for critical accretion. Differentiating Eqs. (\ref{con1s1}) and (\ref{con11}) with respect to $r$, we obtain, respectively,
\begin{eqnarray}
\label{c1}
\frac{1}{u}u'+\frac{1}{n}n' &=& -\frac{2}{r},\\
\label{c2}
 uu'+\left(1-\frac{2M}{r}-\sigma r+u^{2}\right)\frac{a^{2}}{n}n' &=& -\frac{1}{2}\left(\frac{2M}{r^{2}}-\sigma\right),
\end{eqnarray}
where $'=d/dr$. We rewrite these two equations as
\begin{eqnarray}
\label{c3}
u'&=&\frac{N_{1}}{N},\\
\label{c31}
n'&=&-\frac{N_{2}}{N},
\end{eqnarray}
where
\begin{eqnarray}
\label{c4}
N_{1}&=&\frac{1}{n}\left[\frac{2}{r}\left(1-\frac{2M}{r}-\sigma r+u^{2}\right)a^{2}-\frac{1}{2}\left(\frac{2M}{r^{2}}-\sigma\right)\right],\\
\label{c5}
N_{2}&=&\frac{1}{u}\left[\frac{2u^{2}}{r}-\frac{1}{2}\left(\frac{2M}{r^{2}}-\sigma\right)\right],\\
\label{c6}
N&=&\frac{u^{2}-\left(1-\frac{2M}{r}-\sigma r+u^{2}\right)a^{2}}{nu}.
\end{eqnarray}
Assuming that the flow is continual at every point of spacetime, if the denominators on the right hand sides of the equations (\ref{c3}) and (\ref{c31}) are zero at some points, their numerators must also be zero at these points to avoid discontinuities in the flow. The resulting equations determine the critical point
\begin{eqnarray}
\label{V1}
N_{1}=\frac{1}{n_{\rm c}}\left[\frac{2}{r_{\rm c}}\left(1-\frac{2M}{r_{\rm c}}-\sigma r_{\rm c}+u_{\rm c}^{2}\right)a_{\rm c}^{2}-\frac{1}{2}\left(\frac{2M}{r_{\rm c}^{2}}-\sigma\right)\right]=0,
\end{eqnarray}
\begin{eqnarray}
\label{V2}
N_{2}=\frac{1}{u_{\rm c}}\left[\frac{2u_{\rm c}^{2}}{r_{\rm c}}-\frac{1}{2}\left(\frac{2M}{r_{\rm c}^{2}}-\sigma\right)\right]=0,
\end{eqnarray}
and
\begin{eqnarray}
\label{V3}
N=\frac{u_{\rm c}^{2}-\left(1-\frac{2M}{r_{\rm c}}-\sigma r_{\rm c}+u_{\rm c}^{2}\right)a_{\rm c}^{2}}{n_{\rm c}u_{\rm c}}=0,
\end{eqnarray}
where $a_{\rm c}\equiv a(r_{\rm c})$ and $u_{\rm c}=u(r_{\rm c})$ with $r_{\rm c}$ denotes the position of the critical point. From Eqs. (\ref{V1}), (\ref{V2}), and (\ref{V3}), we obtain the radial velocity and the speed of sound at the critical points, respectively,
\begin{eqnarray}
\label{Va}
u_{\rm c}^{2}=\frac{1}{4}\left(\frac{2M}{r_{\rm c}}-\sigma r_{\rm c}\right),
\end{eqnarray}
\begin{eqnarray}
\label{cs}
a_{\rm c}^{2}=\frac{u_{\rm c}^{2}}{1-\frac{2M}{r_{\rm c}}-\sigma r_{\rm c}+u_{\rm c}^{2}}.
\end{eqnarray}
Physically acceptable solutions of Eqs. (\ref{c3}) and (\ref{c31}) exist if $u_{\rm c}^{2}\geq 0$ and $a_{\rm c}^2\geq 0$, therefore we easily get
\begin{eqnarray}
\label{ineq1}
\left\{ \begin{array}{l@{\quad \quad } l}
\displaystyle u_{\rm c}^{2}=\frac{1}{4}\left(\frac{2M}{r_{\rm c}}-\sigma r_{\rm c}\right)\geq0,\\
\displaystyle a_{\rm c}^{2}=\frac{u_{\rm c}^{2}}{1-\frac{2M}{r_{\rm c}}-\sigma r_{\rm c}+u_{\rm c}^{2}}=\frac{\frac{1}{4}\left(\frac{2M}{r_{\rm c}}-\sigma r_{\rm c}\right)}{1-\frac{3M}{2r_{\rm c}}-\frac{5}{4}\sigma r_{\rm c}}\geq0.
\end{array}
\right.
\end{eqnarray}
These equations give the conditions which the critical points must fulfill
\begin{eqnarray}
\label{V11}
\left(1-\sqrt{1-\frac{15}{2}\sigma M}\right)\frac{2}{5\sigma}<r_{\rm c}\leq\sqrt{\frac{2M}{\sigma}}.
\end{eqnarray}
For $\sigma= 1/8M$, this equation gives $12M/5<r_{\rm s}\leq 4M$, meaning the critical points lie in the black hole. Because $\left(1-\sqrt{1-\frac{15}{2}\sigma M}\right)\frac{2}{5\sigma}< r_-<\sqrt{\frac{2M}{\sigma}}<r_+$ for $0<\sigma< 1/8M$, Eq. (\ref{V11}) implies that the critical points lie behind the outer horizon, while the critical points in Schwarzschild spacetime can lie out of the horizon in the black hole \cite{Malec:1999dd}.

\section{The polytropic solution}
In this section, we will consider the accretion rate for polytrope gas and calculate the gas compression or the adiabatic temperature profile at the outer horizon.
\subsection{Accretion for polytrope gas}
We first calculate explicitly $\dot{M}$ by assuming the accretion matter with the polytrope equation of state \cite{Bondi:1952ni, michel1972accretion}
\begin{eqnarray}
\label{p1}
p=Kn^\gamma,
\end{eqnarray}
where the adiabatic index satisfies $1<\gamma<5/3$ and $K$ is a constant. Inserting this equation into the conservation equation of mass-energy (\ref{cme}) and integrating, one can easily get
\begin{eqnarray}
\label{ro1}
\rho=\frac{K}{\gamma-1}n^{\gamma}+mn,
\end{eqnarray}
where $mn$ is the rest energy density with $m$ an integration constant. With the definition of the speed of sound (\ref{ss}), the Bernoulli equation (\ref{con3}) can be rewritten as
\begin{eqnarray}
\label{con31}
\left(1+\frac{a^{2}}{\gamma-1-a^{2}}\right)^{2}\left(1-\frac{2M}{r}-\sigma r+u^{2}\right)=\left(1+\frac{a_{\infty}^{2}}{\gamma-1-a_{\infty}^{2}}\right)^{2}.
\end{eqnarray}
Using the radial velocity (\ref{Va}) and the critical speed of sound (\ref{cs}), this equation should satisfy at the critical points
\begin{eqnarray}
\label{conc}
\left(1-\sqrt{\frac{2}{M\sigma}} a_{\rm c}^{2}\right)\left(1-\frac{a_{\rm c}^{2}}{\gamma-1}\right)^{2}=\left(1-\frac{a_{\infty}^{2}}{\gamma-1}\right)^{2}.
\end{eqnarray}
For $r_{\rm c}\leq r <\infty $, the baryons still are non-relativistic, we can expect $a^2<a_{\rm c}^{2}\ll 1$ and expand (\ref{conc}) to the first order in $a^2_{\rm c}$ and $a^2_{\rm \infty}$ to have
\begin{eqnarray}
\label{arc}
a_{\rm c}^{2}\approx\frac{1}{1+\frac{\gamma-1}{\sqrt{2M\sigma}}} a_{\infty}^{2}.
\end{eqnarray}
We then can express the critical radius $r_{\rm c}$ in terms of the boundary condition $a_{\rm \infty}$ and the black-hole mass $M$ as
\begin{eqnarray}
\label{rc}
r_{\rm c}\approx \left[1-\sqrt{\frac{2}{M\sigma}} a_{\rm c}^{2}\right]\sqrt{\frac{2M}{\sigma}}\approx \left[1-\frac{1}{\sqrt{\frac{M\sigma}{2}}+\frac{\gamma-1}{2}} a_{\infty}^{2}\right]\sqrt{\frac{2M}{\sigma}}.
\end{eqnarray}
From Eqs. (\ref{ss}), (\ref{p1}) and (\ref{ro1}), we get
\begin{eqnarray}
\label{ns}
\gamma Kn^{\gamma-1}=\frac{ma^{2}}{1-a^{2}/(\gamma-1)}.
\end{eqnarray}
Since $a^2\ll 1$, one has $n\sim a^{2/(\gamma-1)}$ and
\begin{eqnarray}
\label{nn}
\frac{n_{\rm c}}{n_{\infty}}\approx\left(\frac{a_{\rm c}}{a_{\infty}}\right)^{\frac{2}{\gamma-1}}.
\end{eqnarray}
Because equation (\ref{acc}) is independent of $r$, it must also hold at the critical points which we can use to determine the Bondi accretion rate
\begin{eqnarray}
\label{rom}
\dot{M}&=&4\pi r^{2}mnu=4\pi r_{\rm c}^{2}mn_{\rm c}u_{\rm c}\\\nonumber
&=&\frac{8\pi M}{\sigma} mn_{\infty}a_{\infty}\left(1-\frac{2}{\sqrt{\frac{M\sigma}{2}}+\frac{\gamma-1}{2}}~a_{\infty}^{2}\right)
\left(1+\frac{\gamma-1}{\sqrt{2M\sigma}}\right)^{-\frac{1}{\gamma-1}}\left(1+\frac{\gamma-1}{\sqrt{2M\sigma}}-\sqrt{\frac{2}{M\sigma}}~a_{\infty}^{2}\right)^{-\frac{1}{2}}.
\end{eqnarray}
It is obvious that the quintessence parameter $\sigma$ plays an important role in the accretion process. For $r_{\rm c}\leq r <\infty $, since $a_{\rm \infty}\ll 1$, the accretion rate (\ref{rom}) reduces to $\dot{M}\simeq \frac{8\pi M}{\sigma} mn_{\infty}a_{\infty} \left(1+\frac{\gamma-1}{\sqrt{2M\sigma}}\right)^{-(\gamma+1)/2(\gamma-1)}$.

\subsection{Asymptotic behavior at the outer horizon}
Near the outer horizon $r= r_{+}$, the fluid velocity can be approximated by
\begin{eqnarray}
\label{vh}
v^{2}\approx\frac{2M}{r}+\sigma r.
\end{eqnarray}
At $r_{+}$, the flow speed approximately equals the speed of light, $v^2(r_{+})\approx 1$. Using equations (\ref{acc}), (\ref{rom}), and (\ref{vh}), we derive the gas compression at the outer horizon,
\begin{eqnarray}
\label{nh}
\frac{n_{\rm h}}{n_{\infty}}=A\left[2M\left(\frac{1+\left(1-8M\sigma\right)^{\frac{1}{2}}}{2\sigma}\right)^{3}+\sigma \left(\frac{1+\left(1-8M\sigma\right)^{\frac{1}{2}}}{2\sigma}\right)^{5}\right]^{-\frac{1}{2}},
\end{eqnarray}
where
\begin{eqnarray}
A=\frac{2M a_{\infty}}{\sigma}\left(1-\frac{2}{\sqrt{\frac{M\sigma}{2}}+\frac{\gamma-1}{2}}a_{\infty}^{2}\right)\left(1+\frac{\gamma-1}{\sqrt{2M\sigma}}\right)^{-\frac{1}{\gamma-1}}
\left(1+\frac{\gamma-1}{\sqrt{2M\sigma}}-\sqrt{\frac{2}{M\sigma}}a_{\infty}^{2}\right)^{-\frac{1}{2}}.
\end{eqnarray}
Using equations (\ref{p1}) and (\ref{nh}), we obtain the adiabatic temperature profile at the outer horizon, assuming a Maxwell-Boltzmann gas $p=nk_{\rm B} T$
\begin{eqnarray}
\label{th}
\frac{T_{\rm h}}{T_{\infty}}=A^{\gamma-1}\left[2M\left(\frac{1+\left(1-8M\sigma\right)^{\frac{1}{2}}}{2\sigma}\right)^{3}+\sigma \left(\frac{1+\left(1-8M\sigma\right)^{\frac{1}{2}}}{2\sigma}\right)^{5}\right]^{-\frac{\gamma-1}{2}}.
\end{eqnarray}
Since $a_{\rm \infty}\ll 1$ for $r_{\rm s}\leq r <\infty $, the parameter $A$ reduces to $A=\frac{2M a_{\infty}}{\sigma}\left(1+\frac{\gamma-1}{\sqrt{2M\sigma}}\right)^{-\frac{\gamma+1}{2(\gamma-1)}}$ and further reduces to $A=\frac{2M a_{\infty}}{\sigma}(2\gamma-1)^{-\frac{\gamma+1}{2(\gamma-1)}}$ for $\sigma=\frac{1}{8M}$, so consequently we have $\frac{n_{\rm h}}{n_{\infty}}=a_{\infty}(2\gamma-1)^{-\frac{\gamma+1}{2(\gamma-1)}}$ and $\frac{T_{\rm h}}{T_{\infty}}=a^{\gamma-1}_{\infty}(2\gamma-1)^{-\frac{\gamma+1}{2}}$.

\section{Conclusions and discussions}
Here we considered accretion onto a Kiselev black hole with static exterior spacetime surrounded by a quintessence field. We obtained the fundamental equations for the flow of matter onto the black hole without the back-reaction. We derived the general analytic expressions for the critical points and the mass accretion rate and determined the physical conditions the critical points should fulfill. For a polytropic gas, we gave the explicit expressions for the mass accretion, the gas compression, the temperature profile at the outer horizon. The results have shown that the accretion rate and the critical points are clearly dependent on the quintessence parameter $\sigma$, which may be an observational feature to incorporate in astrophysical applications.

\begin{acknowledgments}
This study is supported in part by National Natural Science Foundation of China (Grant Nos. 11147028 and 11273010), Hebei Provincial Natural Science Foundation of China (Grant No. A2014201068), the Outstanding Youth Fund of Hebei University (No. 2012JQ02), and the Midwest universities comprehensive strength promotion project.
\end{acknowledgments}

\bibliographystyle{elsarticle-num}
\bibliography{ref}

\end{document}